\documentclass[aps,prl,twocolumn,showpacs,groupedaddress]{revtex4}  
\usepackage{graphicx,subfigure}  
\usepackage{dcolumn}   
\usepackage{bm}        
\usepackage{amssymb}   
\usepackage{amsmath} 
\usepackage[english]{babel}

\newcommand{\J}{G}
\newcommand{\h}{z}
\newcommand{\Tr}{\mbox{Tr}}

\newcommand{\be}{\begin{equation}} 
\newcommand{\ee}{\end{equation}}
\newcommand{\bea}{\begin{eqnarray}}
\newcommand{\eea}{\end{eqnarray}}

\newcommand{\opone}{\leavevmode\hbox{\small1\kern-3.3pt\normalsize1}}
\newcommand{\betaminusone}{(\beta\!-\!1)}
\begin{document}

\title{Adaptive gene regulatory networks}
\author{Franck Stauffer$^1$
and Johannes Berg$^{2}$}
\affiliation{$^1$Institut f\"ur Theoretische Physik,
Universit\"at zu K\"oln\\
Z\"ulpicher Stra{\ss}e 77,
50937 K\"oln, Germany\\
$^2$Physikalisches Institut
Albert-Ludwigs-Universit\"at Freiburg\\
Herrmann Herder-Str. 3, 
79104 Freiburg, Germany}
\date{\today}

\begin{abstract} \noindent Regulatory interactions between genes show
  a large amount of cross-species variability, even when the
  underlying functions are conserved: There are many ways to achieve
  the same function. Here we investigate the ability of regulatory
  networks to reproduce given expression levels within a simple model
  of gene regulation. We find an exponentially large space of
  regulatory networks compatible with a given set of expression
  levels, giving rise to an extensive \textit{entropy of networks}.
  Typical realisations of regulatory networks are found to share a
  bias towards symmetric interactions, in line with empirical
  evidence.
\end{abstract}


\maketitle
\noindent


The expression of genes is regulated such that the right combinations
of gene products are generated at the right time and place of an
organism. Key regulators of gene expression are \textit{transcription
  factors}, proteins which bind to specific sites on DNA and influence
the expression of nearby genes. Typically, the expression of a gene is
effected by a combination of several transcription factors, and conversely, a
transcription factor regulates several genes. Expression levels can
thus depend on the entire set of regulatory interaction between
transcription factors and their target genes, referred to as a
\textit{regulatory network}. These intracellular reaction networks
process extracellular information to induce specific gene expression
patterns, allowing, for instance, the development of a complex body
plan, or responses to external conditions.

Even though regulatory networks are tuned carefully to
produce specific expression patterns, there are in general many 
networks fulfilling a regulatory task. An example is the
control of mating type in different yeast species: The same set of
genes controlled in \textit{S. cerevisiae} by an activator which is
upregulated in a certain state is controlled by a repressor which is
downregulated in that state in \textit{C. albicans}~\cite{Tsongetal:2003}.  A
second prominent example is the development of the anterior patterning
in insect embryos, leading to the formation of the insect's head.  The
gene crucial to this process in the fruit fly \textit{Drosophila}, called bicoid, is
absent in many other insects, where a combination of different genes
take on the same task~\cite{Schroeder:2003}. Even whole sets of genes which 
are co-expressed across the entire yeast family can have different regulatory interactions 
in different species~\cite{Tanayetal:2005}. Source of these changing interactions is 
a rapid evolutionary turnover of transcription factor binding sites at the level of DNA 
sequences~\cite{Tautz:2000,MustonenLassig:2005}. This can generate new
regulatory interactions.
A recent essay on the degeneracy of regulatory networks 
can be found in~\cite{Chouard:2008}.

The large number of regulatory networks with a given function (viable
networks) is particularly relevant from an evolutionary perspective,
as neutral evolution gradually explores different viable networks. 
Viable networks can form a set with intricate geometric features in
the space of regulatory network.  An analogy is the set of all
RNA-sequences which fold into a given secondary structure, which
stretches across the entire sequence space~\cite{Schusteretal:1994}.
Numerical studies, based on simple models of gene
regulation~\cite{BornholdtSneppen:1998} found that the space of viable
networks can be crossed in small steps, and that a wide range of new
expression patterns can be generated by small changes to different
viable
networks~\cite{CilibertiMartinWagner:2007a,CilibertiMartinWagner:2007b}.

\begin{figure}[bt]
\includegraphics[width=0.45\textwidth]{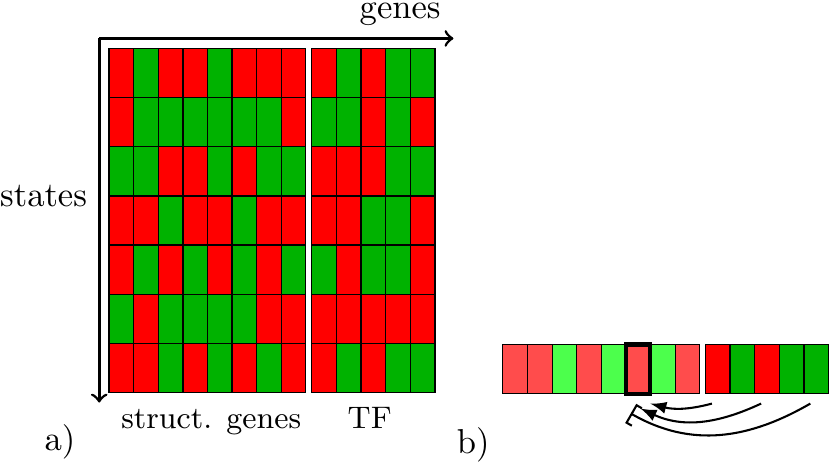}
\caption{\label{fig1}{\bf Expression levels and regulatory networks.}
  a)  We list 
  genes along the $x$-axis, and states of the organism along the
  $y$-axis.  Following established convention in expression analysis, expression levels are colour-coded with high expression
  levels shown in red (dark), low levels in green (light).
  b) Regulatory interactions must be compatible with gene expression 
levels in all states of the organism. The schematic example shows interactions between
  transcription factors and a single target gene; two enhancing
  interactions ($\rightarrow$) with upregulated transcription factors, and
  a repressive interaction ($\dashv$) with a downregulated
  transcription factor lead to the activation of the target gene.
}
\end{figure}

These observations call for a statistical approach based on the
ensemble of all viable networks, which is the topic of this paper.  We
consider a model with two classes of genes: \textit{structural genes}
(coding e.g. for enzymes or cellular components), and
\textit{transcription factors}. The expression levels of structural
genes are prescribed for different states of the organism and are \textit{fixed} for a 
given state. For
instance, when nutrients are available, specific enzymes have to be
produced to digest these nutrients. On the other hand, the expression
levels of transcription factors, and the regulatory interactions between genes 
can be \textit{adapted} to meet the expression levels of structural genes. The
freedom to alter expression levels of transcription factors turns out
to be crucial.

In the following, we develop a simple model based on \textit{quenched} random expression
levels of structural genes, and \textit{adaptive} regulatory
interactions and expression levels of transcription factors. 
The ensemble of viable networks
is characterized by the microcanonical partition function 
\be 
\label{eq:Z}
Z \equiv e^S = \frac{\Tr_{{\bf J}, {\bf\xi}_t } I({\bf J}, {\bf\xi})
}{\Tr_{{\bf J}, {\bf\xi}_t } } \ ,  
\ee 
giving the fraction of viable networks in terms
of the trace $\Tr_{{\bf J}, {\bf\xi}_t }$ over the phase space
(regulatory interactions ${\bf J}$ and the expression levels of
transcription factors ${\bf\xi}_t$) and an indicator function $I({\bf
  J}, {\bf\xi})$ of couplings and all expression levels. The indicator function is 
defined to equal one for a viable network and zero
otherwise.

Specifically, genes are labelled $i=1,\ldots,N$ for structural genes
and $i=N+1,\ldots,\beta N$ for transcription factors. The regulatory
network is encoded in a matrix of regulatory interactions $J_{ij}$,
with positive $J_{ij}$ indicating that gene $j>N$ produces a
transcription factor which enhances the expression of gene $i \neq j$,
and represses gene $i$ for negative values of $J_{ij}$. Different external
and internal states of the organism are labelled by
$\mu=1,\ldots,\alpha N$. $\xi_i^{\mu}$ denotes the (log-)expression
level of gene $i$ in state $\mu$, and is positive for high
concentrations of the gene product and negative for low
concentrations. Assuming transcription factors act independently on
their target genes, the condition for a viable network is modelled as 
\be
\label{viable}
\xi_i^{\mu}/\sqrt{N} \sum_j J_{ij} \xi_j^{\mu} > \kappa \ \ \  \forall i,\mu \ .
\ee
Threshold condition (\ref{viable}) has been used extensively to model
neural~\cite{HertzKroghPalmer} and gene regulatory
networks~\cite{WahdeHertz:2000,BuchlerGerlandHwa:2003,
CilibertiMartinWagner:2007a,CilibertiMartinWagner:2007b}.
The indicator function for a viable network in the partition function~(\ref{eq:Z})
can be written in terms of the Heaviside step-function $\Theta(x)$ as
\bea
\label{eq:indicator}
I(\{ J_{ij},\xi_i^{\mu} \})\!\! &=&\!\! 
\prod_{i,\mu} \Theta\left( \frac{1}{\sqrt{N}} \xi_i^{\mu} 
  \sum_j J_{ij} \xi_j^{\mu} - \kappa \right) \\
&=&\!\! \prod_{i,\mu} \int_{\kappa}^{\infty} \!\!\! d\lambda \int\! \frac{dx}{2\pi}
e^{ i\sum_{i\mu} \lambda_i^{\mu} x_i^{\mu} 
- \frac{i}{\sqrt{N}} \sum_{ij\mu} x_i^{\mu} \xi_i^{\mu} J_{ij} 
\xi_j^{\mu} } \nonumber .
\eea

We constrain vectors of regulatory interactions ${\bf J}_i$ and 
of expression levels ${\bf \xi}_i$ to lie on hyperspheres. This 
defines the trace over phase space~(\ref{eq:trace})
\be
\label{eq:trace}
\Tr = \prod_{i} \int d\mu({\bf J}_i) 
      \prod_{i>N} \int d\mu({\bf \xi}_{i}) 
\ee
with
$d\mu({\bf J}_i) = \prod_{j>N} dJ_{ij} \delta\left( \betaminusone N-
\sum_{j>N} J_{ij}^2 \right)$ and analogously for the 
transcription factor expression levels. The quenched average of~(\ref{eq:Z}) 
over expression levels of structural genes $\langle \langle \ln Z \rangle \rangle=
 \prod_{i\leq N} \int d\mu({\bf \xi}_{i}) \ln Z$ is performed using the replica trick. 

\begin{figure}[h!]
\vspace{-6.75cm}
\includegraphics[width=0.35\textwidth]{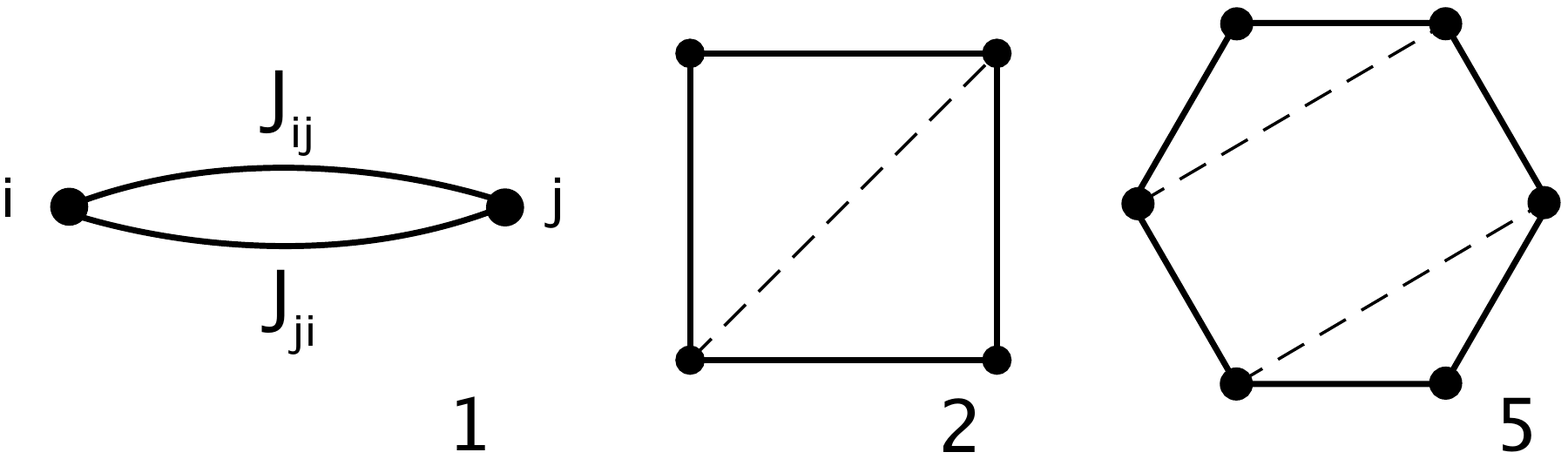}
\put(-12,28){\large$\cdots$}\\
\includegraphics[width=0.35\textwidth]{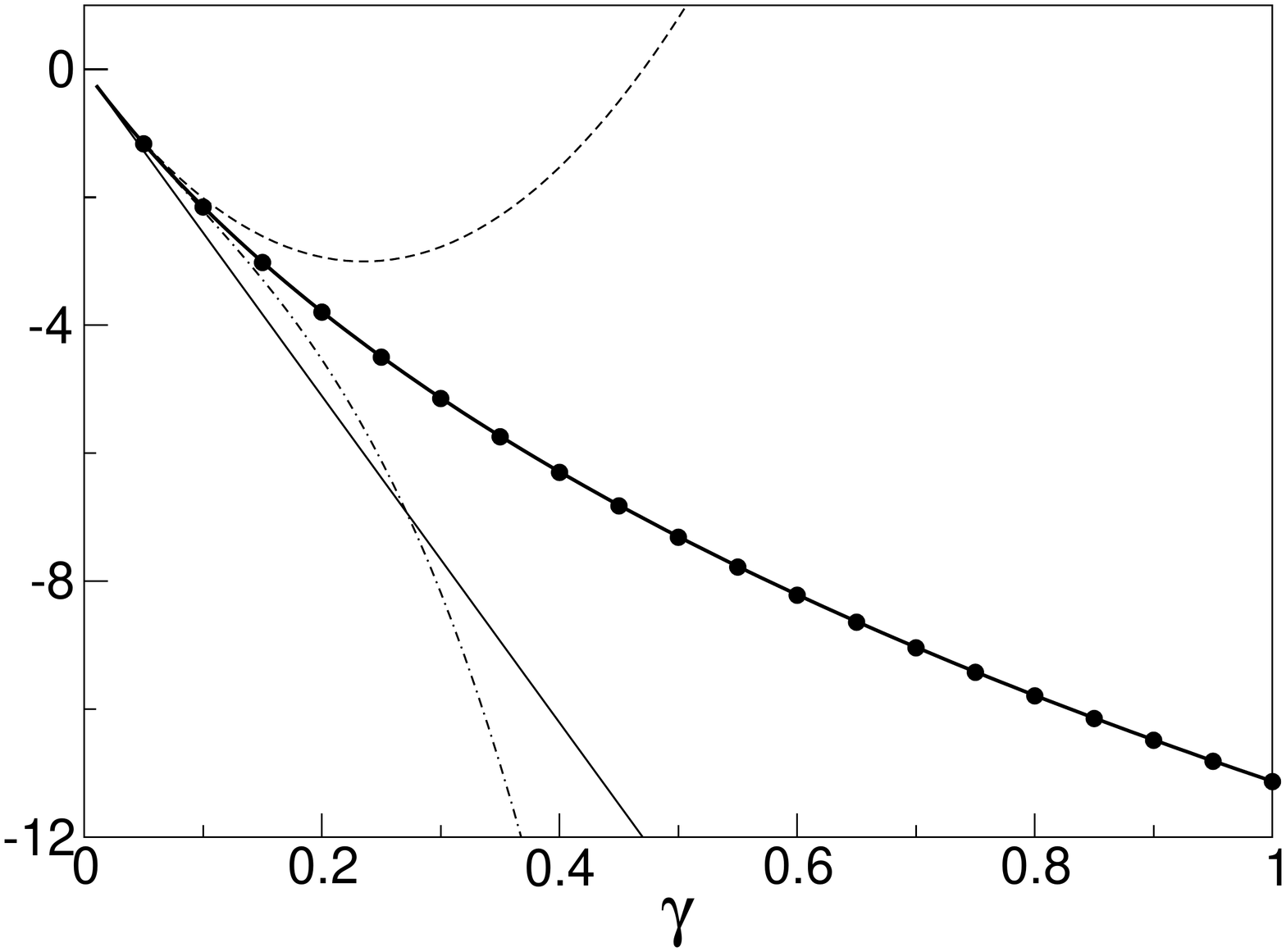}
\caption{\label{fig2} {\bf Averaging over expression levels.}  a) The
  diagrams corresponding to the first three terms in~(\ref{eq:av}) are
  shown along with their combinatorial factors. Nodes represent
  variables $i,j,\ldots$, solid lines indicate the corresponding
  matrix entries $\J_{ij}$, dashed lines are contractions $i=j$.  b)
  Plotting the logarithm of~(\ref{eq:av}) against $\gamma$ shows the
  contribution of different diagrams. The first diagram gives a linear
  term (thin solid line), the series up to second and third order are
  shown by the dashed and dashed-dotted curves respectively. These are
  valid approximations up to some finite values of $\gamma$ only. The thick
  solid line gives the full series to infinite order (thick solid
  line) along with a numerical computation of~(\ref{eq:av}), where
  $\J_{ij}$ was taken a random matrix of size $N=50$ with
  i.i.d. normally distributed elements.}
\end{figure}

Transcription factors play a special role; their expression levels provide the regulatory
input for \textit{every gene} in the regulatory network. This produces
an effective coupling between regulatory interactions of different
genes. One consequence emerges already at the level of the
average of~(\ref{eq:indicator}) over the expression levels $\xi$. As an illustration, we
consider a toy problem, where the average of
$\exp\{-i\sqrt{\gamma/(2 N)} \sum_{ij} \xi_i \J_{ij} \xi_j\}$ is
computed over a distribution of independent normally distributed
variables $\xi_i$. $\J_{ij}$ is a symmetric matrix with uncorrelated random entries.
\bea
\label{eq:av}
&&\langle e^{-\frac{i}{2}\sqrt{\gamma/N} \sum_{ij} \xi_i \J_{ij} \xi_j}\rangle = 
1/\sqrt{ \det( \opone + i\sqrt{\gamma/N} {\bf \J}) } \nonumber \\
&=&\exp\{ -\frac{1}{2} \sum_{n=1}^{\infty} \frac{1}{n} \mbox{tr} (i\sqrt{\gamma/N} {\bf \J})^n \} \\
&=& \exp\{ \gamma/4 \sum_i \h_i - \gamma^2/4 \sum_i \h_i^2 + 5\gamma^3/12 \sum_i \h_i^3 + \cdots \} 
\nonumber\\
&=& \exp\{ \sum_i w( \gamma \h_i ) \} \nonumber
\eea
with shorthand $\h_i=1/N \sum_j \J_{ij} \J_{ji}$ and $\mbox{tr}$ denoting the 
matrix trace. The successive terms in the power series 
(\ref{eq:av}) can be represented diagrammatically; Fig.~\ref{fig2}~a)
shows the first three diagrams. Fig.~\ref{fig2}~b) shows how the different 
powers in~(\ref{eq:av}) contribute to the average and how for finite values of 
$\gamma$ the series has to be taken to infinite order, giving 
$w(\h)=\frac{2 \h +1 - \sqrt{1+4 \h} }{ 8 \h } - 
\frac{1}{2} \log \left( \frac{1}{2}+\frac{1}{2}\sqrt{1+4\h} \right)$. 
This is in contrast to the standard situation in fully-connected disordered 
models, where in the thermodynamic limit the series in~(\ref{eq:av})
terminates after the first term.

The approach~(\ref{eq:av}) applied to the full  model~(\ref{eq:Z})-(\ref{eq:indicator}) 
gives $\sum_{i \leq N,\mu} w(\h_i^{\mu}) + \sum_{i > N,\mu,a} w(\h_i^{\mu a})$ with $\h_i^{\mu}= \sum_a (x_i^{\mu a})^2$ and  
$\h_i^{\mu a}=  (x_i^{\mu a})^2 + \frac{1}{N} \sum_j  (x_i^{\mu a})^2 (J_{ji}^{a})^2 + \frac{2}{N} \sum_j x_i^{\mu a} x_j^{\mu a}J_{ij}^{a} J_{ji}^{a}$. 
Neglecting fluctuations of $\h_i^{\mu a}$ across genes, 
the entropy of viable networks $\langle\langle S\rangle\rangle \equiv N^2 s$
can be computed in the thermodynamic limit $N\to \infty$ by
standard methods. Within a replica-symmetric ansatz we obtain
\begin{eqnarray}
\label{eq:entropy}
\!\!\!\!\!&&s = \mbox{extr} \left[
    \frac{1}{2} \alpha\betaminusone\!\left(F-\ln F\right) + \frac{1}{2} \alpha\hat{X}_2^{-}X_2^{-} \right.  \\
&&\left.+ \frac{1}{2}\alpha\betaminusone\hat{X}_2^{+}X_2^{+}       +\alpha\betaminusone\hat{X}_1X_1+\frac{\betaminusone^2}{4}\ln\left(1-h^2\right) \right.\nonumber \\
  && \left. + \alpha \betaminusone \ln\left[
      \int_\kappa^\infty d\lambda\int \frac{dx}{2\pi} 
      \exp \left\{ w\left(\frac{1}{F^2}\left[x^2  + X_2^{+}   
\right.\right.\right.\right.\right.\nonumber \\
  &&\left.\left.\left.\left.\left.  -2i xh X_1  \right] 
          + \frac{X_2^{-}}{\betaminusone F} \right) + i(\lambda-\hat{X}_1)x  -\frac{1}{2}\hat{X}_2^{+} x^2 \right\} \right] \right. \nonumber\\
  && \left.+ \alpha \int d\hat{y} dy e^{\left[iy\hat{y}+w\left(\frac{y}{F}\right) \right]}\ln\left[ H\left(\frac{\kappa}{\sqrt{\hat{X}_2^{-}+2iy}}\right)\right] 
  \right] \nonumber \ .
\end{eqnarray}
The entropy $s$ of viable networks is determined by the extremum over the saddle
point parameters $F,X_1,\hat{X}_1,X_2^{\pm},\hat{X}_2^{\pm}$, and $h$. 
Saddle point parameter $h$ has an intuitive interpretation in terms 
of the symmetry of regulatory interactions and will be discussed below.
$H(x)$ denotes the cumulative Gaussian measure $\int_x^{\infty}\!\!
\frac{dy}{\sqrt{2\pi}} \exp\{-y^2/2\}$.

\begin{figure}[tbh]
\includegraphics[width=0.35\textwidth]{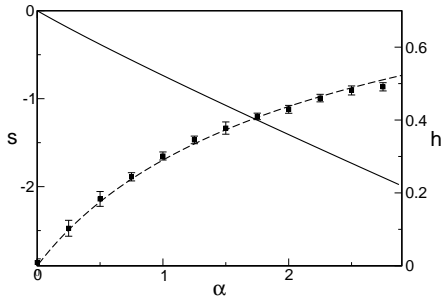}
\caption{\label{fig3}{\bf Entropy and symmetry of viable networks.}
With increasing number of patterns $P=\alpha N$, the space of viable networks shrinks, 
and the networks become increasingly symmetric, see text. Here the entropy $s$ per 
structural gene (solid line) and symmetry parameter
$h$ (dashed line) are plotted against $\alpha$ for $\beta=2,\kappa=0$. The
{\tiny $\blacksquare$}-symbols stem from numerical simulations with $N=80$,
averaged over $20$ realizations of the quenched disorder (mean and
standard error). The numerics is based on simulated annealing under
Monte-Carlo dynamics of the regulatory interactions $J_{ij}$ and the
expression levels of transcription factors $\xi^{\mu}_{i>N}$.
}
\end{figure}

The entropy of viable networks~(\ref{eq:entropy}) decreases with increasing number of
patterns $P=\alpha N$, see Fig.~\ref{fig3}. This is to be expected, as
each set of expression patterns induces a new set of constraints on
the network. However, the entropy remains finite
even as the number of patterns becomes large with $\alpha \to \infty$:
there (typically) always exists a viable network, and 
there is no transition to a phase where solutions of (\ref{viable}) no
longer exist. Such phase transitions are well known in neural networks
and many combinatorial problems~\cite{HertzKroghPalmer}. In contrast, the ability
of regulatory networks to store a large number
of expression patterns of structural genes stems directly from the
freedom to choose expression levels of transcription factors:
transcription factor expression levels adapt in such a way that
regulatory interactions compatible with expression levels of all genes
can be found.

The saddle point parameter $h=\frac{1}{\betaminusone^2 N^2}\sum_{i,j}
J_{ij}J_{ji}$ is the \textit{symmetry parameter} of the resulting regulatory
network. A positive value of $h$ indicates that if gene
$i$ regulates gene $j$, and also $j$ regulates $i$, the signs of these
interactions are correlated, with like signs occurring more
frequently than opposite signs. The origin of this symmetry lies in
condition (\ref{viable}) for a viable network, where a positive
value of $\xi^{\mu}_i \xi^{\mu}_j$ for some $i,j,\mu$ gives rise to
positive values for both $J_{ij}$ \textit{and} $J_{ji}$, and
analogously for negative values.  Thus the symmetry parameter $h$
increases with the number of expression patterns;
Fig.~\ref{fig3} shows the analytical result for $h$ along with the outcome of
numerical simulations.

This statistical bias towards symmetric interactions is compatible
with empirical data on regulatory networks. A literature search for  
well-documented cases of mutually interacting genes with known 
interaction sign
finds $9$ cases of mutually interacting gene pairs with like
interaction
sign~\cite{symmetric}
compared to only $3$ cases with different
sign~\cite{asymmetric}.
A nontrivial statistics of reciprocal interactions has also been found in
neural and metabolic networks~\cite{GarlaschelliLoffredo:2004}, where,
however, the signs of the interactions are generally unknown.

\begin{figure}[tb!]
\includegraphics[width=0.35\textwidth]{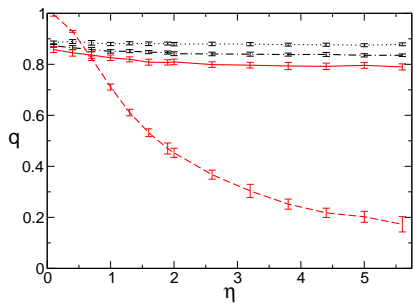}
\caption{\label{fig4}{\bf Response to changing expression levels.}
  The overlaps of perturbed and unperturbed systems (see text) are
  plotted against the perturbation strength $\eta$: structural genes
  expression level overlaps (red dashed line) tend to zero with
  increasing $\eta$ by construction, whereas TF expression level
  overlaps (red solid line) quickly reach a plateau. The same holds
  for regulatory interactions to TF (black dotted line), and
  interactions to structural genes (black dash-dotted). The plateau
  value decreases with the fraction $1-\beta$ of TF in the genome. The
  data stem from Monte-Carlo simulations with $N=40$, $\alpha=1$,
  $\beta=2$, and $\kappa=0$, averaged over $20$ samples.}
\end{figure}

Over long evolutionary timescales, the required expression levels of
structural genes can change. In the case of enzymes, for instance,
changing nutrient availability or changing metabolic rates alter the
required expression levels. Such changes of the expression levels of
structural genes induce adaptive changes both of the regulatory
network, and of the expression levels of transcription factors. To
investigate the adaptation to changing expression levels of structural
genes, we systematically perturb the expression levels of structural
genes of a viable network, rendering it, in general, at first
unviable. (Expression levels $\xi_{i \leq N}^{\mu}$ are perturbed by
adding i.i.d. Gaussian random variables with mean zero and standard
deviation $\eta$ and normalizing their variances to one again.)
Subsequently, regulatory interactions and transcription factor
expression levels are adapted until the viability
condition~(\ref{viable}) is satisfied again. The overlap $q_{\xi}^<=
\frac{1}{NP}\sum_{i\leq N,\mu} \xi_i^{\mu} \xi_i^{\prime\mu}$ of
structural gene expression levels of the unperturbed (unprimed) and
the perturbed (primed) system quantifies the strength of the
perturbation, the analogously defined $q_{\xi}^>,q_{J}^<,q_{J}^>$
quantify the response of the system to this
perturbation. Figure~\ref{fig4} shows the overlaps as a function of
perturbation strength. One finds that already small perturbations with
$q_{\xi}^< \approx 1$ result in a drop of the overlaps to a plateau
value. Larger perturbations, and even the limit $q_{\xi}^< \to 0$
induce only a slow decay of $q_{\xi}^>,q_{J}^<,q_{J}^>$ from their
plateau values.  Accordingly, close to any viable network for one set
of expression levels of structural genes, there exists a viable
network for any other, even unrelated set of expression levels. This
effect allows fast adaptation to changes in the required expression
levels.

Another consequence of the observed drop of the TF expression level
overlap to a plateau is that expression levels of TF change
\textit{more} than those of structural genes for small
perturbations. For large perturbations, the expression levels of TF
change \textit{less} than those of structural genes. This effect may
explain an apparent contradiction in the cross-species comparison of
experimentally measured expression levels. A comparison of humans with
other primates shows large changes of TF expression levels~\cite{Giladetal:2006}
compared to structural genes, different \textit{Drosophila}
species show only small changes of TF expression levels compared to
structural genes~\cite{RifkinKimWhite:2003}.

In summary, we have investigated the degeneracy of regulatory networks
within a simple model of genetic regulation. Whereas the connection between
annealed TF expression levels and the large space of viable networks
is likely to persist also in more complex models, the geometry of this
space may well change. In particular, models taking into account
physical interactions between transcription factors to implement
logical functions~\cite{BuchlerGerlandHwa:2003} lead to p-spin
interactions $J_{ijk\ldots}$ and may result in a disconnected solution
space and combinatorial complexity.

\begin{acknowledgments}
Many thanks to A. Altland and M. Cosentino Lagomarsino for discussions.
Funding from the DFG is acknowledged under 
grant BE 2478/2-1 and SFB 680.
\end{acknowledgments}


\end{document}